\documentclass[preprint,superscriptaddress,amsmath,amssymb,prd,aps,showpacs,floatfix,nofootinbib]{revtex4-1}

\usepackage[T1]{fontenc}
\usepackage{lmodern}

\usepackage[dvipsnames,x11names]{xcolor}

\usepackage[colorlinks=true,linkcolor=Blue,citecolor=ForestGreen,urlcolor=NavyBlue]{hyperref}

\usepackage{graphicx}
\usepackage[sort&compress]{natbib}
\usepackage{lipsum}
\usepackage{morefloats}
\usepackage[pdf]{pstricks}
\usepackage{subfigure}
\usepackage{amsmath}
\usepackage{amssymb}
\usepackage{amsfonts}
\usepackage{rotating}
\usepackage{cancel}
\usepackage{mathtools}
\usepackage{bbm}
\usepackage{dsfont}
\usepackage{bbold}
\usepackage{multirow}
\usepackage{ulem}
\usepackage{physics}
\usepackage{orcidlink}

\newcommand{\mev}{\textrm{ MeV}}

\begin{document}

\frenchspacing

\title{\boldmath $\bar B^0 \to \bar K^{(*) \,0} X$, $B^- \to K^{(*) \, -} X$, $\bar B_s^0 \to \eta (\eta', \phi) X$ 
from the $X(3872)$ molecular perspective}

\author{Wei-Hong Liang\orcidlink{0000-0001-5847-2498}}%
\email{liangwh@gxnu.edu.cn}
\affiliation{Department of Physics, Guangxi Normal University, Guilin 541004, China}%
\affiliation{Guangxi Key Laboratory of Nuclear Physics and Technology, Guangxi Normal University, Guilin 541004, China}%

\author{Ting Ban}%
\affiliation{Department of Physics, Guangxi Normal University, Guilin 541004, China}%

\author{Eulogio Oset\orcidlink{0000-0002-4462-7919}}%
\email{Eulogio.Oset@ific.uv.es}
\affiliation{Department of Physics, Guangxi Normal University, Guilin 541004, China}%
\affiliation{Departamento de F\'{\i}sica Te\'orica and IFIC, Centro Mixto Universidad de
Valencia-CSIC Institutos de Investigaci\'on de Paterna, Aptdo.22085,
46071 Valencia, Spain}%

\begin{abstract}
  We study the decays $\bar B^0 \to \bar K^0 \, X$, $B^- \to K^- \, X$, $\bar B^0_s \to \eta (\eta')\, X$, $\bar B^0 \to \bar K^{*0} \, X$, $B^- \to K^{*-} \, X$ , $\bar B_s^0 \to \phi \, X$, with $X \equiv X(3872)$, from the perspective of the $X(3872)$ being a molecular state made from the interaction of the $D^{*+} D^-, D^{*0} \bar D^0$ and $c.c.$ components. We consider both the external and internal emission decay mechanisms and find an explanation for the $\bar K^0 \, X$ and $K^- \, X$ production rates, based on the mass difference of the charged and neutral $D^* \bar D$ components. We also find that the internal and external emission mechanisms add constructively in the $\bar B^0 \to \bar K^0 \, X$, $B^- \to K^- \, X$ reactions, while they add destructively in the case of $\bar B^0 \to \bar K^{*0} \, X$, $B^- \to K^{*-} \, X$ reactions. This feature explains the decay widths of the present measurements and allows us to make predictions for the unmeasured modes of $\bar B^0_s \to \eta (\eta')\, X(3872)$ and $B^- \to K^{*-} \, X(3872)$. The future measurement of these decay modes will help us get a better perspective on the nature of the $X(3872)$ and the mechanisms present in production reactions of that state.

\end{abstract}

\maketitle

\section{Introduction}
The $X(3872)$, as the first state found demanding an exotic, non $q\bar q$, nature,
has been the subject of intense debate concerning its structure.
Some groups claim a molecular nature with the $D^{*+} D^-$, $D^{*0} \bar D^0$ and $c.c.$ components \cite{Braaten:2020nmc,Liu:2020tqy,Meng:2020cbk,Dong:2021juy,Wu:2021udi,Gordillo:2021bra,Dong:2021bvy,Meng:2021kmi,Kamiya:2022thy,Lin:2022wmj,
Ji:2022uie,Wang:2022qxe,Wang:2022xga,Kinugawa:2023fbf,Yang:2023mov,Wu:2023rrp,Peng:2023lfw,Terashima:2023tun},
other groups claim a compact tetraquark state \cite{Shi:2021jyr,Esposito:2021vhu,Huang:2021poj,Chen:2022ddj,Sharma:2022ena, Wang:2013vex},
or some mixture of the two structures \cite{Lebed:2022vks,Wang:2023ovj} (see Refs.~\cite{Kalashnikova:2018vkv,Yamaguchi:2019vea,Brambilla:2019esw,Guo:2019twa,Chen:2022asf,Mai:2022eur} for earlier related work).
In Ref.~\cite{Kang:2016jxw}, the analysis of spectra in $X(3872)$ production is inconclusive concerning the compositeness of the state.
The extreme proximity of the state to the $D^{*0} \bar D^0$ threshold has been a main factor to support the molecular nature \cite{Guo:2017jvc,Dong:2021juy},
but, as discussed in detail in Ref.~\cite{Song:2023pdq},
the proximity of the state to a threshold is not sufficient to guarantee its molecular nature, although certainly makes it far more likely.
It is found in Ref.~\cite{Song:2023pdq}
that it is possible to have a state of nonmolecular nature very close to a threshold,
but one pays a huge price:
the scattering length $a$ for the hadron-hadron components,
to which the state inevitably couples,
goes to zero,
and the effective range $r_0$ goes to infinity in the limit of zero energy binding.
With the small but finite binging of the $X(3872)$, the resulting $a, r_0$ magnitudes if one demands the state to be of nonmolecular nature are very small and very large,
respectively, compared with experimental data, to the point that one can exclude the nonmolecular nature with uncertainties smaller than about $5\%$.

Very recently, in the work of Ref.~\cite{ShenWang},
the $\bar B^0 \to \bar K^0 \, X(3872)$ and $B^- \to K^- \, X(3872)$ reactions were proposed as a means to learn about the nature of the $X(3872)$.
Indeed, with a compact tetraquark nature for the $X(3872)$,
one expects the branching ratio of the two decay modes to be equal,
but experimentally $\mathcal{B}(B^- \to K^- \, X(3872))/\mathcal{B}(\bar B^0 \to \bar K^0 \, X(3872)) \simeq 2$.
Instead, from the molecular perspective this ratio is tied to the loop functions $G_i$ for the neutral and charged components $D^{*0 }\bar D^0$, $D^{*+} D^-$, respectively,
which are different at the pole of the state because of the mass difference between the charged $D^{*+} D^-$ and neutral $D^{*0}\bar D^0$ components.
The study of Ref.~\cite{ShenWang} gave a natural explanation for this experimental ratio based on the molecular nature of the $X(3872)$ state.

In Ref.~\cite{ShenWang}, the decay from the molecular perspective is studied considering the dominant external emission decay mode.
In the present work we follow closely the idea of Ref.~\cite{ShenWang} but include in the study also the internal emission.
Then extend the idea to study the
$\bar B^0 \to \bar K^{*0} \, X(3872)$, $B^- \to K^{*-} \, X(3872)$ and $\bar B^0_s \to \phi \, X(3872)$ decays.
The latter reaction proceeds only via internal emission, and we would expect a small rate compared to that of the $\bar B^0 \to \bar K^{*0} \, X(3872)$ decay,
but, surprisingly, the rates are similar.
Its explanation is found, because in the case of the $\bar K^0 , K^-$ pseudoscalar production the external and internal emission mechanisms add constructively,
while in the case of $\bar K^{*0}, K^{*-}$ production they add destructively.
We compare ratios of branching ratios for presently measured decay modes and make predictions for the unmeasured modes,
$\bar B^0_s \to \eta (\eta') \, X(3872)$ and $B^- \to K^{*-} \, X(3872)$.
It will be interesting to compare these predictions with future measurements,
which will certainly help to go deeper in our understanding of the nature of the $X(3872)$ and the mechanisms of its production.

\section{Formalism}

\subsection{The molecular $X(3872)$ state}
We follow here the work of Ref.~\cite{Song:2023pdq}, where the $X(3872)$ is obtained from the interaction of the $D^{*0} \bar D^0, D^{*+} D^-$ ($+c.c.$) components.
The extended local hidden gauge approach is used,
with the coupled channels
\begin{equation}\label{eq:XcXn}
  \begin{split}
    X_{\rm c} =& \dfrac{1}{\sqrt{2}} (D^{*+} D^- -D^{*-} D^+), \\[2mm]
   X_{\rm n} =&  \dfrac{1}{\sqrt{2}} (D^{*0} \bar D^0 -\bar D^{*0} D^0).
  \end{split}
\end{equation}
The interaction between these components stems from the exchange of $\rho$ and $\omega$ vector mesons,
and with the label $X_{\rm c} \equiv 1, X_{\rm n}\equiv 2$,
one finds the interaction matrix
\begin{align}\label{eq:V}
\centering
V=
\left(
  \begin{array}{cc}
    \tilde{v}  & \tilde{v}\\
    \tilde{v} & \tilde{v}\\
  \end{array}
\right),
\end{align}
with
\begin{align}\label{eq:V2}
\tilde{v} =-g^2\;\dfrac{4 \, m_{D^{*0}}\,m_{D^0}}{M^2_V};  ~~~~ g=\dfrac{M_V}{2\, f}; ~~~ M_V =800\, \mev,~~~ f=93\, \mev,
\end{align}
which has been calculated at the $D^{*0} \bar D^0$ threshold.
The isospin states, with the isospin multiplets phase convention $(D^+, -D^0)$, $(\bar D^0, D^-)$ (and same for $D^*$),
are given by
\begin{equation}\label{eq:Xisospin}
  \begin{split}
   |X, I=0\rangle  = \dfrac{1}{\sqrt{2}} \, (X_{\rm c} +X_{\rm n}), \\[2mm]
   |X, I=1, I_3=0\rangle = \dfrac{1}{\sqrt{2}} \, (X_{\rm c} -X_{\rm n}),
  \end{split}
\end{equation}
and we see that for $I=0$ the interaction is attractive and can produce a bound state,
but for $I=1$ the interaction is zero and we do not expect a bound state.
We note in passing that the strength of the interaction here is double than for $I=0$ in the $T_{cc}$ state \cite{FeijooLiang}.
This means we should expect a bigger binding for the $X(3872)$ than for the $T_{cc}$.
With the centroid of mass data and the $X(3872)$ mass of the PDG \cite{pdg},
which we write below,
\begin{alignat}{2}\label{eq:masses}
& m_{{D}^0} = (1864.84 \pm 0.05) \, \mev,        &\quad \quad &m_{{D}^+} = (1869.66 \pm 0.05) \, \mev, \nonumber\\[1mm]
& m_{D^{*0}} = (2006.85 \pm 0.05) \, \mev, &\quad \quad &m_{D^{*+}} = (2010.26 \pm 0.05) \, \mev, \\[1mm]
&m_{D^{*0}}+m_{{D}^0} =3871.69 \, \mev , &\quad \quad & m_{D^{*+}} + m_{{D}^-} =3879.92 \, \mev, \nonumber\\[1mm]
&M_{X(3872)} =(3871.65 \pm 0.07) \, \mev, \nonumber
\end{alignat}
the binding of the $X(3872)$ with respect to the $D^{*0} \bar D^0$ threshold is of $0.04\, \mev$,
but can be larger considering the uncertainties of the masses.

The scattering matrix with this potential is easily obtained as
\begin{align}\label{eq:T}
\centering
T=[1-VG]^{-1}\, V = \dfrac{1}{\rm det}\;
\left(
  \begin{array}{cc}
    \tilde{v}  & \tilde{v}\\
    \tilde{v} & \tilde{v}\\
  \end{array}
\right),
\end{align}
with det being the determinant of $[1-VG]$, given by
\begin{equation}\label{eq:det}
  {\rm det} = 1- \tilde{v} \, G_1 - \tilde{v} \, G_2,
\end{equation}
and the diagonal $D^* \bar D$ loop function $G= {\rm diag} [G_1, G_2]$ given with the cutoff regularization by

\begin{equation}\label{eq:G}
  G_i = \int_{|\vec q\,| < q_{\rm max}} \dfrac{{\rm d}^3 q}{(2\pi)^3} \; \dfrac{\omega_1^{(i)}(\vec q\,)+\omega_2^{(i)}(\vec q\,)}{2\,\omega_1^{(i)}(\vec q\,)\,\omega_2^{(i)}(\vec q\,)}\; \dfrac{1}{s-[\omega_1^{(i)}(\vec q\,)+\omega_2^{(i)}(\vec q\,)]^2 +i \epsilon},
\end{equation}
with $\omega_j^{(i)}(\vec q\,)=\sqrt{m_j^2 + \vec q^{\; 2}}$ the energy of particle $j$ in the $i$th channel.
We use $q_{\rm max}= 420 \, \mev$, as demanded in the study of the $T_{cc}$,
with $D^* D$ components \cite{FeijooLiang},
but to get the nominal binging of $40 \, \rm keV$ we have to multiply $\tilde{v}$ by a factor $\beta =0.537$.
Then we need to know the couplings of the $X(3872)$ to any of the charged or neutral components and we find
\begin{equation}\label{eq:couplings}
  g_1^2 = \lim_{s \to s_0} (s-s_0)\, T_{11}, ~~~~~ g_1 g_2 = \lim_{s \to s_0} (s-s_0)\, T_{12},
\end{equation}
with $s_0$ the squared of the energy of the $X(3872)$ state.
From Eqs.~\eqref{eq:T} and \eqref{eq:det} using l'Hospital rule for the limit,
we find
\begin{equation}\label{eq:g1}
  g_1^2 = \dfrac{\tilde{v}}{-\tilde{v} \frac{\partial}{\partial s}\, (G_1+G_2)} =g_1\, g_2.
\end{equation}
This means that $g_1 =g_2$ (or $g_{\rm c} = g_{\rm n}$ in the nomenclature for charged or neutral components).
In the formalism below we just need $G_1, G_2$ at $s_0$ and the ratio $g_{\rm n}/ g_{\rm c}=1$,
which indicates that we have an $I=0$ state.

In order to get the $40\; \rm keV$ binding, we had to multiply the potential of Eq.~\eqref{eq:V2} by $\beta =0.537$. Since the uncertainty of the $X(3872)$ mass is about $70\; \rm keV$, we can accept a binding of about $110\; \rm keV$ with respect to the $D^{*0} \bar D^0$ threshold. 
If we play with the uncertainties in the $D^0$ and $\bar D^{*0}$ masses, the binding can be of $210 \; \rm keV$ with respect to this threshold,
which means we could have a value of $\beta$ closer to $1$ to get the right binding. Yet, the theory is always accompanied by a cutoff to regularize the loops and this is fine tuned in each case to obtain the desired binding. One should also note that the potential used in Eq.~\eqref{eq:V2} has omitted the range of the vector exchange, replacing $(\vec q^{\, 2}+ M_V^2)^{-1}$ by $(M_V^2)^{-1}$ in the vector exchange propagator.
Consideration of this factor is done in Ref.~\cite{SakaiRoca} and also weakens the strength of the interaction. All these ingredients should be considered to get the proper binding. We should also note that the couplings change moderately with the binding energy $B$ since one has $g^2 \sim \sqrt{B}$ \cite{Weinberg,Baru,Gamermann2}. 
Yet, we refrain from doing such exercises here because the only information that we need, is the ratio of the couplings to $D^0 \bar D^{*0}$ and $D^+ D^{*-}$ and this ratio is unity in our theory, independent on the precise potential and cutoff used.

The formalism that we use here follows closely the one of Ref.~\cite{ShenWang}
to study the $\bar B^0 \to \bar K^{0} \, X(3872)$ and $B^- \to K^{-} \, X(3872)$ decays.
In addition we include the hadronization from the internal emission mode for the $D\bar D^*, \bar D D^*$ components and extend the formalism to the production of vector mesons.

\subsection{Internal emission for pseudoscalar production}
From the molecular point of view, the $X(3872)$ is given by
\begin{equation}\label{eq:WaveFuncX}
  X=\dfrac{1}{\sqrt{2}}(X_{\rm c}+X_{\rm n}),
\end{equation}
where $X_{\rm c}$ and $X_{\rm n}$ are the charged and neutral components of $D \bar D^*, \bar D D^*$ respectively, shown in Eq.~\eqref{eq:XcXn}.
Altogether
\begin{equation}\label{eq:WaveFuncX2}
  X=\dfrac{1}{2}(D^{*+} D^- -D^{*-} D^+ +D^{*0} \bar D^0 -\bar D^{*0} D^0),
\end{equation}
which has isospin $I=0$ and $C$-parity $C=+$.
In $s$-wave this state corresponds to $J^P=1^+$, completing the quantum numbers of the state.

\begin{figure}[b]
\begin{center}
\includegraphics[scale=0.43]{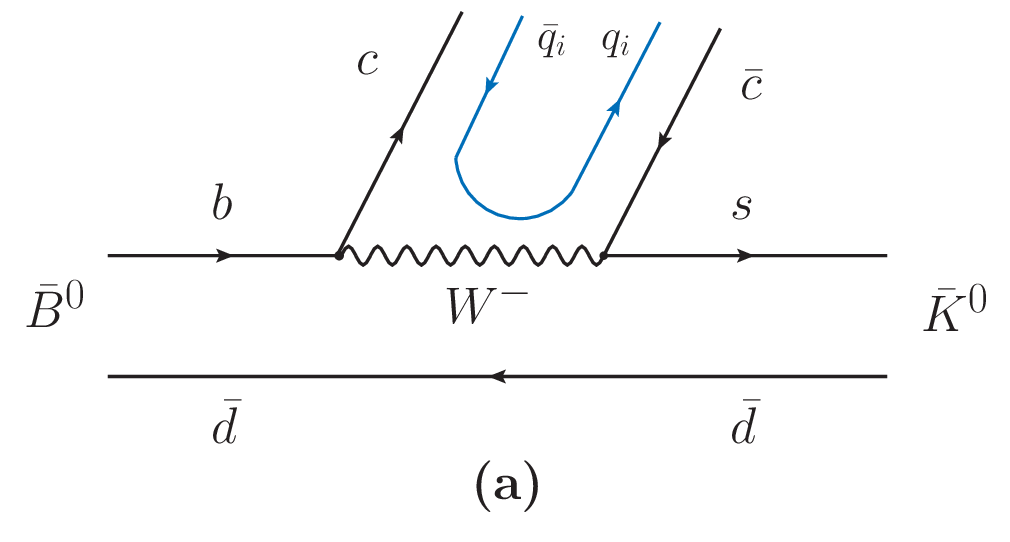}~~~~~
\includegraphics[scale=0.43]{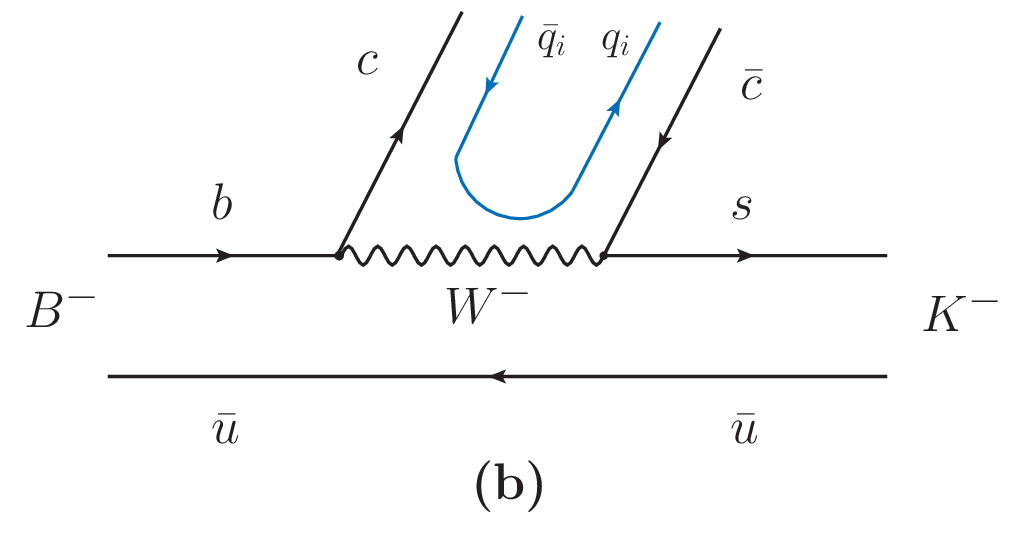}
\end{center}
\vspace{-0.9cm}
\caption{Internal emission mechanism at the quark level for $\bar B^0 \to \bar K^0 \, c\bar c$ decay (a) and $B^- \to K^- \, c\bar c$ decay (b).}
\label{Fig:Fig1}
\end{figure}

At the quark level, the weak decay for $\bar B^0 \to \bar K^0 \, X(3872)$ and $B^- \to K^- \, X(3872)$ proceeds as shown in Fig.~\ref{Fig:Fig1}.
The $X(3872)$ production from these mechanism is accomplished by means of hadronization of the $c\bar c$ component.
Let us see how this proceeds.
We start by writing the pseudoscalar ($P$) and vector ($V$) matrices for $q_i \bar q_j$ in terms of the physical mesons.
With the $\eta, \eta'$ mixing of Ref.~\cite{Bramon},
we have
\begin{equation}\label{eq:Pmatrix}
   P =
    \left(
    \begin{array}{cccc}
    \frac{1}{\sqrt{2}}\pi^0 + \frac{1}{\sqrt{3}}\eta + \frac{1}{\sqrt{6}}\eta' & \pi^+ & K^+  &  \bar D^0\\[2mm]
    \pi^- & -\frac{1}{\sqrt{2}}\pi^0 + \frac{1}{\sqrt{3}}\eta + \frac{1}{\sqrt{6}}\eta' & K^0  & D^-\\[2mm]
    K^- & \bar{K}^0 & ~-\frac{1}{\sqrt{3}}\eta + \sqrt{\frac{2}{3}}\eta'~  & D_s^-\\[2mm]
    D^0 & D^+ & D_s^+  & \eta_c\\
    \end{array}
    \right),
\end{equation}

\begin{equation}\label{eq:Vmatrix}
    V =
    \left(
    \begin{array}{cccc}
    \frac{1}{\sqrt{2}}\rho^0 + \frac{1}{\sqrt{2}}\omega  & \rho^+ & K^{*+}  &  \bar D^{*0}\\[2mm]
    \rho^- & -\frac{1}{\sqrt{2}}\rho^0 + \frac{1}{\sqrt{2}}\omega  & ~K^{*0}~  & D^{*-}\\[2mm]
    K^{*-} & \bar{K}^{*0} & \phi  & D_s^{*-}\\[2mm]
    D^{*0} & D^{*+} & D_s^{*+}  & J/\psi \\
    \end{array}
    \right).
\end{equation}
Then, upon hadronization of the $c\bar c$, we need to create a vector and a pseudoscalar,
but the order matters and we can have $PV$ or $VP$ combinations.
Hence
\begin{equation}\label{eq:PV}
{\rm a)}~PV:~~c\bar c \to \sum_i c \,\bar q_i q_i \,\bar c =  \sum_i P_{4i}\, V_{i4} = (PV)_{44}
     = D^0 \bar D^{*0} + D^+ D^{*-} \,+ \cdots ,
\end{equation}
\begin{equation}\label{eq:VP}
{\rm b)}~VP:~~c\bar c \to \sum_i c \, \bar q_i q_i \, \bar c =  \sum_i V_{4i}\, P_{i4} = (VP)_{44}
     = D^{*0} \bar D^0 + D^{*+} D^- \,+ \cdots,
\end{equation}
%
where the notation ``$\cdots$'' indicates other terms that have no overlap with the $X(3872)$ components of Eq.~\eqref{eq:WaveFuncX2}.
We see that the combination $(VP)_{44} -(PV)_{44}$,
\begin{equation}\label{eq:VP2}
  (VP)_{44} -(PV)_{44}=D^{*+} D^- +D^{*0} \bar D^0 -D^{*-} D^+  -\bar D^{*0} D^0,
\end{equation}
has a perfect match with Eq.~\eqref{eq:WaveFuncX2}.
This hadronization is common to $\bar B^0 \to \bar K^0 \, X(3872)$ and $B^- \to K^- \, X(3872)$.

\begin{figure}[t]
\begin{center}
\includegraphics[scale=0.43]{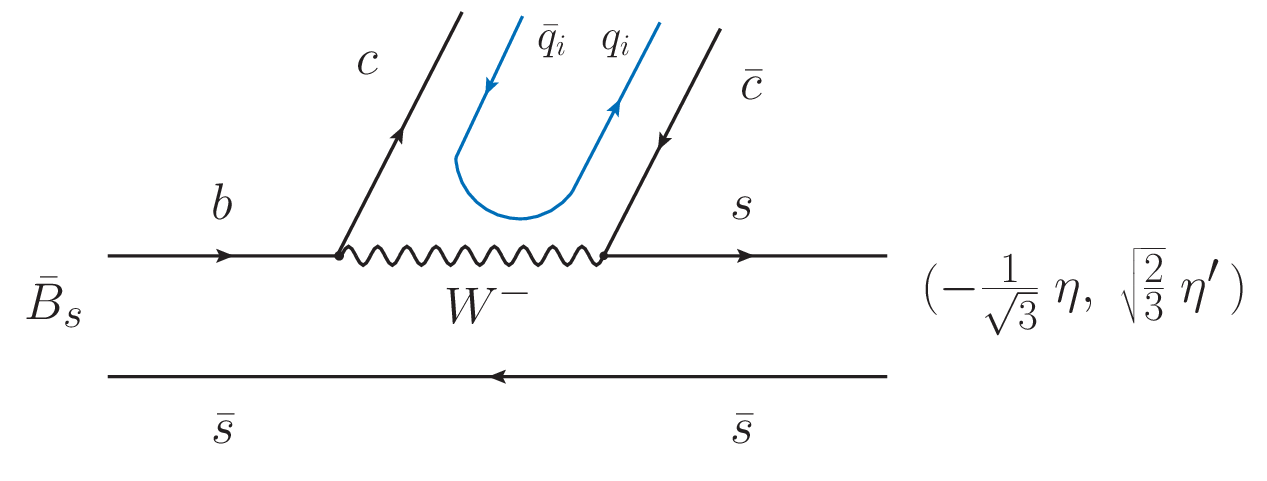}
\end{center}
\vspace{-0.9cm}
\caption{Internal emission mechanism at the quark level for $\bar B_s^0 \to \eta (\eta') \, c\bar c$ decay.}
\label{Fig:Fig2}
\end{figure}

Let us see with the same mechanism, how we can have the decay $\bar B_s^0 \to \eta (\eta') \,X(3872)$.
The mechanism is shown in Fig.~\ref{Fig:Fig2}.
The hadronization in this case gives us the same combination as in Eq.~\eqref{eq:VP2},
and the $s\bar s$ splits into $\eta$ and $\eta'$ according to the matrix element $P_{33}$ of Eq.~\eqref{eq:Pmatrix},
as
\begin{equation}\label{eq:P33}
  s\bar s \to -\frac{1}{\sqrt{3}}\, \eta + \sqrt{\frac{2}{3}}\, \eta'.
\end{equation}

\subsection{External emission for pseudoscalar production}
\begin{figure}[t]
\begin{center}
\includegraphics[scale=0.45]{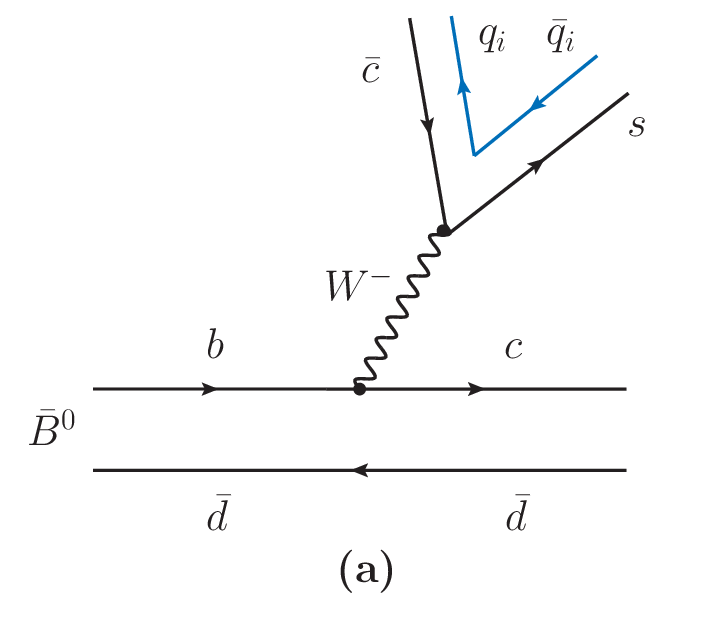}~~~~~~~
\includegraphics[scale=0.45]{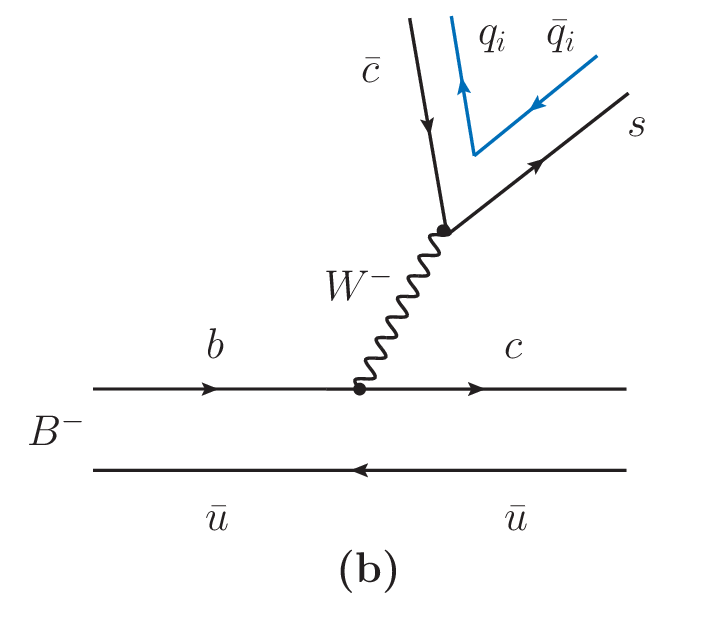}
\end{center}
\vspace{-0.9cm}
\caption{External emission mechanism for $\bar B^0 \to \bar c s \, c\bar d$ decay (a) and $B^- \to \bar c s \, c\bar u$ decay (b).}
\label{Fig:Fig3}
\end{figure}

External emission is the dominant mode of weak decay \cite{Chau} since it is color favored.
One expects something like a factor $3$ times bigger amplitude than for internal emission.
The mechanism for external emission is depicted in Fig.~\ref{Fig:Fig3}.
It is clear that the mechanisms of Figs. \ref{Fig:Fig2} and \ref{Fig:Fig3} are different, since the hadronization is done in different quark pairs, $c\bar c$ in Fig. \ref{Fig:Fig2} and $s\bar c$ in Fig.~\ref{Fig:Fig3}.

Upon hadronization, we have three possibilities,
\begin{itemize}
  \item[a)] $\bar c s \to PV,~ c\bar d \to P \;(D^+)$:
    \begin{equation}\label{eq:PVa}
      s\bar c \to \sum_i s \,\bar q_i q_i \,\bar c = \sum_i P_{3i}\, V_{i4} = (PV)_{34}.
    \end{equation}
  Hence,
  \begin{equation}\label{eq:cscd}
    \bar c s \, c \bar d \to (K^- D^{*0} + \bar K^0 D^{*-} + \cdots) \,D^+.
  \end{equation}
  \item[b)] $\bar c s \to VP,~ c\bar d \to P \;(D^+)$:
    \begin{equation}\label{eq:PVb}
      s\bar c \to \sum_i s \,\bar q_i q_i \,\bar c = \sum_i V_{3i}\, P_{i4} = (VP)_{34}.
    \end{equation}
  Hence,
  \begin{equation}\label{eq:cscd}
    \bar c s \, c \bar d \to (K^{*-} \bar D^0 + \bar K^{*0} D^- + \cdots) \,D^+.
  \end{equation}
  \item[c)] $\bar c s \to PP,~ c\bar d \to V \;(D^{*+})$:
    \begin{equation}\label{eq:PVb}
      s\bar c \to \sum_i s \,\bar q_i q_i \,\bar c = \sum_i P_{3i}\, P_{i4} = (PP)_{34}.
    \end{equation}
  Hence,
  \begin{equation}\label{eq:cscd}
    \bar c s \, c \bar d \to (K^{-} \bar D^0 + \bar K^{0} D^- + \cdots)\, D^{*+}.
  \end{equation}
\end{itemize}
We can see that the combination $(PP)_{34} \,D^{*+} -(PV)_{34} \,D^+$,
\begin{equation}\label{eq:VP7}
  (PP)_{34} \, D^{*+} -(PV)_{34} \,D^+=(D^{*+} D^- -D^{*-} D^+)\, \bar K^0,
\end{equation}
has again perfect overlap with the wave function of Eq.~\eqref{eq:WaveFuncX2}.

We can repeat the procedure for $K^-$ production of Fig.~\ref{Fig:Fig3}(b) and we find again
\begin{equation}\label{eq:VP8}
  (PP)_{34} \, D^{*0} -(PV)_{34} \,D^0=(D^{*0} \bar D^0 - \bar D^{*0} D^0)\, \bar K^-,
\end{equation}
which overlaps with Eq.~\eqref{eq:WaveFuncX2} in the same way.

We should also note that with this decay mode we find no terms that go to $\eta (\eta')\, X(3872)$.
Hence, $\eta, \eta'$ production proceeds only via internal emission.

\begin{figure}[b]
\begin{center}
\includegraphics[scale=0.45]{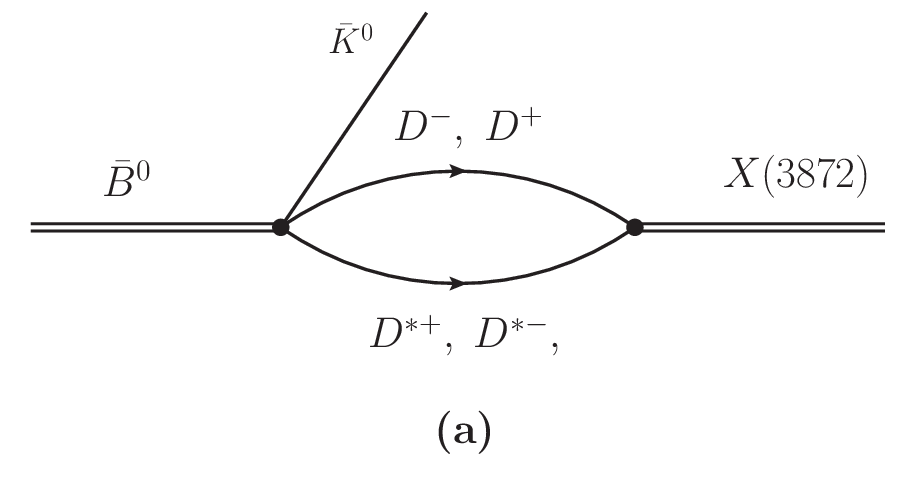}~~~~~
\includegraphics[scale=0.45]{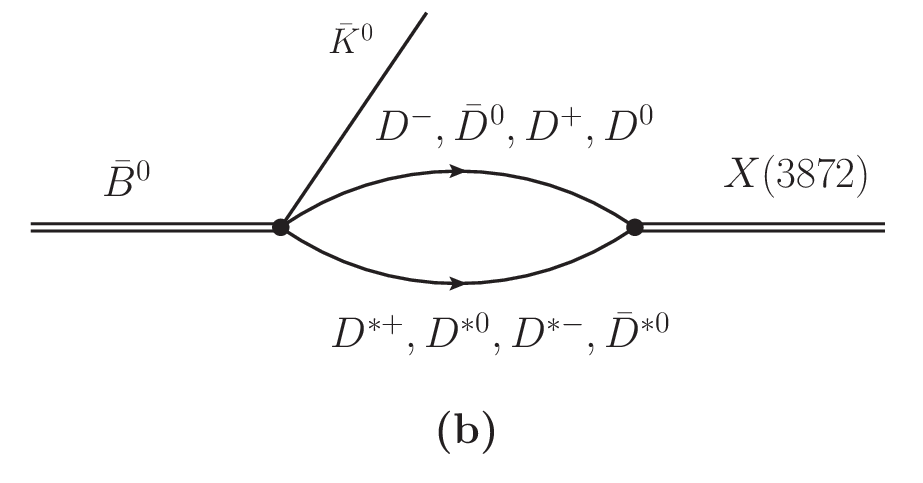}
\end{center}
\vspace{-0.9cm}
\caption{Propagation of $D\bar D^*, \bar D D^*$ components to create the $X(3872)$: (a) external emission; (b) internal emission.}
\label{Fig:Fig4}
\end{figure}

So far we have done the direct production of the components of the $X(3872)$.
In order to produce the resonance, these components must propagate as depicted in Fig.~\ref{Fig:Fig4},
where the loops stand for the $G$ function of $D \bar D^* +c.c.$ propagation.
Then we obtain the amplitude for $\bar B^0 \to \bar K^0 \, X(3872)$ as
\begin{eqnarray}\label{eq:tK0}
  t(\bar K^0)&=& C\, G_{D^{*+}D^-} (M_X)\, g_{\rm c} +A\, \left[ G_{D^{*+} D^-} (M_X)\, g_{\rm c} + G_{D^{*0}\bar D^0} (M_X)\, g_{\rm n} \right] \nonumber\\[2mm]
  &=& C\, g_{\rm c}\, \left\{ G_{D^{*+} D^-} (M_X) + A' \left[G_{D^{*+} D^-} (M_X) + \frac{g_{\rm n}}{g_{\rm c}}\, G_{D^{*0}\bar D^0} (M_X)\right] \right\},
\end{eqnarray}
where $A'\equiv A/C$ and $g_{\rm c}, g_{\rm n}$ are the couplings of $(D^{*+} D^- - D^{*-} D^+)$ and $(D^{*0} \bar D^0 - \bar D^{*0} D^0)$ to $X(3872)$ of Eq.~\eqref{eq:WaveFuncX2}, respectively.

Similarly, for $K^- (\eta, \eta')$ production we find
\begin{eqnarray}\label{eq:tK2}
  t(K^-)&=& C\, g_{\rm c}\, \left\{ \frac{g_{\rm n}}{g_{\rm c}}\, G_{D^{*0} \bar D^0} (M_X) + A' \left[G_{D^{*+} D^-} (M_X) + \frac{g_{\rm n}}{g_{\rm c}}\, G_{D^{*0}\bar D^0} (M_X)\right] \right\}, \label{eq:tK-}\\[2.5mm]
  t(\eta)&=& C\, g_{\rm c}\, A' \left[G_{D^{*+} D^-} (M_X) + \frac{g_{\rm n}}{g_{\rm c}}\, G_{D^{*0}\bar D^0} (M_X)\right] \,(-\frac{1}{\sqrt{3}}), \label{eq:teta} \\[2.5mm]
  t(\eta')&=& C\, g_{\rm c}\, A' \left[G_{D^{*+} D^-} (M_X) + \frac{g_{\rm n}}{g_{\rm c}}\, G_{D^{*0}\bar D^0} (M_X)\right] \,(\sqrt{\frac{2}{3}}). \label{eq:tetap}
\end{eqnarray}
Experimentally $\Gamma (K^-)/ \Gamma (K^0) \sim 2$ \cite{pdg} and now we can see qualitatively the reason for this factor,
because at the pole $(G_{D^{*0}\bar D^0} / G_{D^{*+} D^-})^2 \sim 2$,
as can be seen in the work of Ref.~\cite{Gamermann},
due to the mass difference between the $D^{*+} D^-$ and $D^{*0} \bar D^0$ components.
Assuming the contribution of internal emission small,
this could give a qualitative explanation of the experimental ratio,
which is the idea of Ref.~\cite{ShenWang}.
We aim at being more quantitative here and also extend the formalism to the production of vector mesons that we address below.

Once our formalism is set, it is interesting to show the differences with the formalism of Ref.~\cite{ShenWang}. In this latter work, the mechanism of Fig.~\ref{Fig:Fig2} leads to the $X(3872)$ ``with or without a pair of sea quarks" (see Fig. 1 of Ref.~\cite{ShenWang}), and the amplitude is taken as a contact term
\begin{equation}\label{eq:newa}
  t_1(B^0) =g_{BKX} \, p^\mu_{B^0} \, \epsilon_\mu^* (X);~~~~~
  t_1(B^+) =g_{BKX} \, p^\mu_{B^+} \, \epsilon_\mu^* (X).
\end{equation}
In our formalism, where the $X(3872)$ is assumed to be a molecular state, instead of a contact term of Eq.~\eqref{eq:newa}, we have the term with $A'$ in Eqs.~\eqref{eq:tK0} and \eqref{eq:tK-}, with the explicit propagation of the $D^0 \bar D^{*0}, D^+ D^{*-}+c.c.$ components.
However, it is easy to see that the two procedures are identical. Indeed, the $A'$ terms in Eqs.~\eqref{eq:tK0} and \eqref{eq:tK2} accounting for internal emission, are the same. One might think that because the explicit $G$ functions in those expressions, not accounted for by the operators in Eq.~\eqref{eq:newa}, the expressions could be different, but this is not the case because the $G$ functions are calculated at $M_{\rm inv}=M_X$ and, hence, are constants. Thus, as rightly stated in Ref.~\cite{ShenWang}, ``the hadronized components of $c\bar c$ can be absorbed into the coupling strength of $g_{BKX}$''.

In the present work, we stick to only the molecular components. Furthermore, we take advantage of the experimental findings of the $T_{cc}$ that allow us to choose a $q_{\rm max}$ value to regularize the analogous loops in the $X(3872)$ generation, and then we sum coherently the internal and external emission. In as much as the internal emission term is small, we find that the pictures of Ref.~\cite{ShenWang} and the present one are the same, and so are the conclusions: The experimental ratio of $K^-$ to $\bar K^0$ production favors the molecular structure of the $X(3872)$.

The basic new element that we have in this work is the calculation of $\bar K^{*0}$ and $K^{*-}$ vector production, that is addressed in the next subsections, which we study within the same formalism. The simultaneous explanation of pseudoscalar and vector production gives further support to the molecular picture of the $X(3872)$.

\subsection{Internal emission for vector production}
We now look at the reactions $\bar B^0 \to \bar K^{*0}\, X(3872)$, $B^- \to K^{*-}\, X(3872)$, $\bar B_s^0 \to \phi \, X(3872)$.
The internal emission for vector production is identical to the one exposed before in Figs.~\ref{Fig:Fig1}(a) and \ref{Fig:Fig1}(b),
replacing $\bar K^0$ by $\bar K^{*0}$ and $K^-$ by $K^{*-}$ respectively.
For $\phi$ production, it is also easy since the mechanism is the one of Fig.~\ref{Fig:Fig2},
where now $s\bar s \to \phi$.
The weight $A$ will now be $\tilde{A}$, since the coupling of the $q\bar q$ component to pseudoscalars or vectors have different weights \cite{LiangOset}.

\subsection{External emission for vector production}
Following the steps of the former subsection for external emission we find now
\begin{itemize}
  \item[a)] $\bar K^{*0}$ production:
    \begin{equation}\label{eq:PVKstar0}
       (VV)_{34} \,D^+ - (VP)_{34} \,D^{*+}   ~\to ~(-1) (D^{*+} D^- - D^{*-} D^+)\, \bar K^{*0};
    \end{equation}
 \item[b)] $K^{*-}$ production:
    \begin{equation}\label{eq:PVKstar-}
      (VV)_{34} \, D^0 - (VP)_{34}\, D^{*0}  ~\to ~(-1) (D^{*0} \bar D^0 - \bar D^{*0} D^0)\, K^{*-}.
    \end{equation}
\end{itemize}
Note that we obtain the same expression as in Eqs.~\eqref{eq:VP7} and \eqref{eq:VP8},
simply replacing $\bar K^0$ by $\bar K^{*0}$ and $K^-$ by $K^{*-}$ respectively.
There is only one difference which is the sign.
This different sign will be of relevance for the different rates obtained for pseudoscalar and vector production,
as we shall see.

All this said, the amplitudes for vector production are now
\begin{eqnarray}
  t(\bar K^{*0})&=& \tilde{C}\, g_{\rm c}\, \left\{ - G_{D^{*+} D^-} (M_X) + \tilde{A}' \left[G_{D^{*+} D^-} (M_X) + \frac{g_{\rm n}}{g_{\rm c}}\, G_{D^{*0}\bar D^0} (M_X)\right] \right\}, \label{eq:tKstar0}\\[2.5mm]
  t(K^{*-})&=&  \tilde{C}\, g_{\rm c}\, \left\{ -\frac{g_{\rm n}}{g_{\rm c}}\,G_{D^{*0}\bar D^0} (M_X)
           +\tilde{A}' \left[G_{D^{*+} D^-} (M_X) + \frac{g_{\rm n}}{g_{\rm c}}\, G_{D^{*0}\bar D^0} (M_X)\right] \right\}, \label{eq:Kstar-}\\[2.5mm]
  t(\phi)&=& \tilde{C}\, g_{\rm c}\, \tilde{A}' \left[G_{D^{*+} D^-} (M_X) + \frac{g_{\rm n}}{g_{\rm c}}\, G_{D^{*0}\bar D^0} (M_X)\right], \label{eq:tphi}
\end{eqnarray}
where $\tilde{A}'\equiv \tilde{A}/\tilde{C}$.

Note that in Eqs.~\eqref{eq:tK0} to \eqref{eq:tetap} and here in Eqs.~\eqref{eq:tKstar0} to \eqref{eq:tphi}
the constants $\tilde{C}$ and $C$ are different,
hence we cannot relate pseudoscalar production and vector production,
but we can compare different rates for pseudoscalar production and different rates for vector production.
There is one more thing, while $C$ and $\tilde{C}$ and $A, \tilde{A}$ are different,
the ratio $A' \equiv A/C$ and $\tilde{A}' \equiv \tilde{A}/\tilde{C}$ indicates the fraction of internal to external emission,
and this is basically related to the color factor.
Hence, we shall make the reasonable assumption that $A' \equiv \tilde{A}' $.

In addition, for angular momentum conservation we have the extra factor $p_{B_{(s)}}^\mu \, \epsilon_\mu (X)$ for pseudoscalar decay \cite{ShenWang}
and $\epsilon_\mu (V) \, \epsilon^\mu (X)$ for vector decay.
Upon squaring the $t_i$ amplitudes, $|t_i|^2$, and summing over polarizations,
the first term for pseudoscalar production gives $\vec p_X^{\;2} \, M_{B_{(s)}}^2 / M_X^2$ and the second term for vector production a constant.
Since $|\vec p_X| \equiv | \vec p_P|$ in the rest frame of $B_{(s)}$, with $\vec p_P$ the pseudoscalar momentum, for calculating ratios
we just include the $\vec p_P^{\;2}$ factor in the formula of the width.

\section{Results}
In Eqs.~\eqref{eq:tK0} to \eqref{eq:tphi} we have the amplitudes for each one of the transitions considered.
The decay width corresponding to these amplitudes are
\begin{equation}\label{eq:width}
  \Gamma_{i\to j} = \dfrac{1}{8\pi}\; \dfrac{1}{M^2_{B_i}} \; |t_{i, j}|^2 \, p_j \, F(p_j),
\end{equation}
with
\begin{equation}\label{eq:pj}
  p_j= \dfrac{\lambda^{1/2}(M^2_{B_i}, m^2_j, M^2_X)}{2\, M_{B_i}},
\end{equation}
where $B_i$ is the decaying $B_{(s)}$ meson and $j$ the pseudoscalar or vector produced in addition to the $X(3872)$.
The factor $F(p_j)$, where a constant factor, inoperative in ratios, is omitted is
\begin{align}\label{eq:F}
   F(p_j)=\left\{
   \begin{array}{ll}
   p_j^2,  & {\rm for ~ pseudoscalar ~ production}; \\[1mm]
   1,      & {\rm for ~ vector ~ production}.\\
   \end{array}
\right.
\end{align}

When calculating ratios between pseudoscalar or vector production rates,
the factors $C\, g_{\rm c}$ and $\tilde{C}\, g_{\rm c}$ cancel and the ratios depend on the only parameter $A' \equiv \tilde{A}'$.
The present experimental situation, according to the PDG \cite{pdg},
is the following:
\begin{eqnarray}\label{eq:ExpRatio}
  \mathcal{B}(\bar B^0 \to \bar K^0 \, X(3872)) &=& (1.1 \pm 0.4)\times 10^{-4},  \nonumber \\[1.5mm]
  \mathcal{B}(B^- \to K^- \, X(3872)) &=& (2.1 \pm 0.7)\times 10^{-4},  \nonumber \\[1.5mm]
  \mathcal{B}(\bar B^0 \to \bar K^{*0} \, X(3872)) &=& (1.0 \pm 0.5)\times 10^{-4},  \\[1.5mm]
  \mathcal{B}(B^- \to K^{*-} \, X(3872)) &<& 6 \times 10^{-4},  \nonumber \\[1.5mm]
  \mathcal{B}(\bar B_s^0 \to \phi \, X(3872)) &=& (1.1 \pm 0.4)\times 10^{-4}.\nonumber
\end{eqnarray}
We determine the value of $A'$ to get the ratio between $\mathcal{B}(\bar B^0_s \to \phi \, X)/ \mathcal{B}(\bar B^0 \to \bar K^{*0} \, X)$
and then the rest of the ratios are predictions.
We find
\begin{equation}\label{eq:A'}
  A'=0.2,
\end{equation}
which makes sense as the color suppressed internal emission versus external emission modes.
Then we find the ratios of Tables \ref{tab:tab1} and \ref{tab:tab2}.
\begin{table}[b]
	\caption{Ratios of $R\equiv \mathcal{B}(B_{i} \to j)/ \mathcal{B}(\bar B^0 \to \bar K^0 \, X(3872))$. The different life times $\tau_{B_i}$ are considered. In brackets experimental ratio with errors summed in quadrature.}
\centering \vspace{0.3cm}
\begin{tabular*}{0.6\textwidth}{@{\extracolsep{\fill}}r| c c c}
\toprule
 & $B^- \to K^- \, X$  &  $\bar B_s^0 \to \eta \, X$  &  $\bar B_s^0 \to \eta' \, X$  \\
\hline
~~$R$~~   & $2.07~~(1.91\pm 0.94)$  &  $0.043$  &  $0.048$   \\
\hline\hline
\end{tabular*}
\label{tab:tab1}
\end{table}
\begin{table}[h]
	\caption{Ratios of $R' \equiv \mathcal{B}(B_{i} \to j)/ \mathcal{B}(\bar B^0 \to \bar K^{*0} \, X(3872))$. The different life times $\tau_{B_i}$ are considered. In brackets experimental ratio with errors summed in quadrature.}
\centering \vspace{0.3cm}
\begin{tabular*}{0.5\textwidth}{@{\extracolsep{\fill}}r| c c }
\toprule
 & $B^- \to K^{*-} \, X$  &  $\bar B_s^0 \to \phi \, X$    \\
\hline
~~$R'$~~   & $5.2~~(< 6)$  &  $1.10~~(1.10\pm 0.68)$    \\
\hline\hline
\end{tabular*}
\label{tab:tab2}
\end{table}

\begin{figure}[t]
\begin{center}
\includegraphics[scale=0.6]{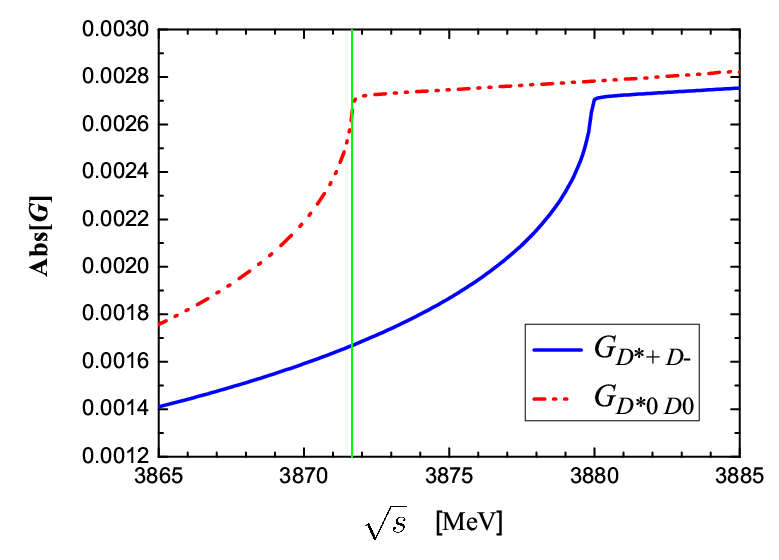}~~~~~~
\includegraphics[scale=0.6]{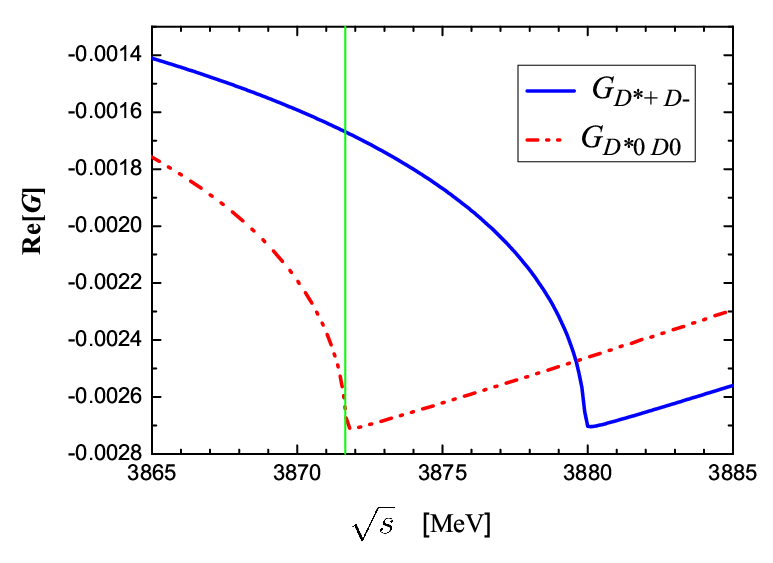}
\end{center}
\vspace{-0.9cm}
\caption{$G_{D^{*+} D^-}$ and $G_{D^{*0} \bar D^0}$ as a function of the energy $\sqrt{s}$. The vertical line indicates the mass of the $X(3872)$.}
\label{Fig:Fig5}
\end{figure}

The essential ingredient in understanding these results is the fact that $|G_{D^{*0} \bar D^0}|$ at the pole is bigger than $|G_{D^{*+} D^-}|$.
We see that $G_{D^{*+} D^-}$ looks like $G_{D^{*0} \bar D^0}$ but displaced about $7\; \mev$ to the right,
and both are negative.
This has a consequence that $|G_{D^{*0} \bar D^0}| > |G_{D^{*+} D^-}|$ at the pole as can be seen in Fig.~\ref{Fig:Fig5}.

It is interesting to remark that the contribution of the internal emission,
even if small, has been important to get these results.
Indeed, if we neglect the internal emission then $\mathcal{B}(B^- \to K^- X)/ \mathcal{B}(\bar B^0 \to \bar K^0 \, X)=2.5$,
which is bigger than experiment.
On the other hand, in this extreme case $\mathcal{B}(\bar B_s^0 \to \phi X)=0$.
The introduction of the internal emission term,
adding constructively to the external emission in Eqs.~\eqref{eq:tK-} to \eqref{eq:tetap}, softens the former ratio.
On the other hand, for vector production in Eqs.~\eqref{eq:tKstar0} and \eqref{eq:Kstar-},
the external and internal emissions interfere destructively for $\bar K^{*0}$ and $K^{*-}$ production, reducing their widths,
and has a consequence that the rates of $\bar B^0 \to \bar K^{*0} \, X(3872)$ and $\bar B_s^0 \to \phi \, X(3872)$ are similar.

\section{Conclusions}
   In this work we have addressed the issue of $B$ decays to a pseudoscalar and the $X(3872)$ and a vector and $X(3872)$,
   from the perspective that the $X(3872)$ is a dynamically generated state from the $D \bar D^* + c.c.$ interaction in coupled channels.
   In particular we have studied the decays 
   $\bar B^0 \to \bar K^0 \, X$, $B^- \to K^- \, X$, $\bar B^0_s \to \eta (\eta')\, X$, $\bar B^0 \to \bar K^{*0} \, X$, $B^- \to K^{*-} \, X$ , $\bar B_s^0 \to \phi \, X$, with $X \equiv X(3872)$.
   We have seen that the production of the final states requires both the mechanism of external and internal emission.
   Although the internal emission mechanism is suppressed with respect to external emission by a color factor,
   we have seen that for the vector production the two terms interfere destructively,
   while they add constructively for pseudoscalar production.
   This has as a consequence that while $\eta$ and $\eta'$ production,
   which proceed via internal emission,
   are small compared to $\bar K^0$ of $K^-$ production,
   which proceeds via the sum of the two mechanisms, $\phi$ production,
   which also proceeds via internal emission, has comparable strength to $K^*$ production,
   where both external and internal emission are at work but add destructively.
   The reason for an unexpected factor $2$ between $K^-$ and $\bar K^0$ production is traced back to the composite nature of the $X(3972)$ from
   the charged and neutral $X_{\rm c}, X_{\rm n}$ of Eq.~\eqref{eq:XcXn} and the mass difference between $D^{*+}\, D^-$ and $D^{*0} \bar D^0$, the same reason exposed in Ref.~\cite{ShenWang}.
Indeed, the production of the $X(3872)$ requires the propagation of the $D^* \bar D$ components, which is accomplished by means of the loop $G$ functions.
The one of the charged components is displaced about $7 \, \mev$ to higher energies,
   due to the mass difference between the charged and neutral components,
   and this has as a consequence that at the pole energy the ratio $|G_{\rm n}/G_{\rm c}|^2$ is about $2.5$ (see Fig.~\ref{Fig:Fig5}).
   This would be the ratio between $K^-$ and $\bar K^0$ production rates in the absence of internal emission.
   The contribution of internal emission reduces this factor to about $2$.
   This provides a natural explanation for this apparently strange ratio that naively one would expect to be of the order of unity.
   The fact that the rates of vector production also appear fine with the same framework adds extra support to this picture.
   In addition we have made predictions for $\eta$ and $\eta'$ production in the pseudoscalar sector,
   and $K^{*-}$ in the vector sector.
   Future measurements of these decays widths should provide extra support for this picture.

\begin{acknowledgments}
We acknowledge useful discussions with J.J. Xie concerning the work of Ref.~\cite{ShenWang}.
This work is partly supported by the National Natural Science Foundation of China (NSFC) under Grants No. 11975083 and No. 12365019, and by the Central Government Guidance Funds for Local Scientific and Technological Development, China (No. Guike ZY22096024).
This project has received funding from the European Union Horizon 2020 research and innovation programme under the program H2020-INFRAIA-2018-1, grant agreement No. 824093 of the STRONG-2020 project.
\end{acknowledgments}


\end{document}